\documentclass[aps,prl,nofootinbib,superscriptaddress,longbibliography]{revtex4-1}

\usepackage{graphicx}
\DeclareGraphicsExtensions{.pdf}

\usepackage{amsfonts,amssymb,amsmath}
\usepackage{amsthm}
\usepackage{tikz}
\usetikzlibrary{calc}
\usetikzlibrary{arrows}
\usepackage{tikz-3dplot}
\usepackage{color} 
\usepackage{graphicx}

\newcommand{\comment}[1]{}

\theoremstyle{plain}
\newtheorem{theorem}{Theorem}

\theoremstyle{definition}

\begin{document}

\title{A Quantum Paradox of Choice: More Freedom Makes Summoning a
  Quantum State Harder}

\author{Emily \surname{Adlam}} \affiliation{Centre for Quantum
  Information and Foundations, DAMTP, Centre for Mathematical
  Sciences, University of Cambridge, Wilberforce Road, Cambridge, CB3
  0WA, U.K.}  \author{Adrian \surname{Kent}} \affiliation{Centre for
  Quantum Information and Foundations, DAMTP, Centre for Mathematical
  Sciences, University of Cambridge, Wilberforce Road, Cambridge, CB3
  0WA, U.K.}  \affiliation{Perimeter Institute for Theoretical
  Physics, 31 Caroline Street North, Waterloo, ON N2L 2Y5, Canada.}

\date{September 2015}

\begin{abstract}
  The properties of quantum information in space-time can be
  investigated by studying operational tasks.  In one such task,
  summoning, an unknown quantum state is supplied at one point, and a
  call is made at another for it to be returned at a third.
  Hayden-May recently proved necessary and sufficient conditions for
  guaranteeing successful return of a summoned state for finite sets
  of call and return points when there is a guarantee of at most one
  summons.  We prove necessary and sufficient conditions when there
  may be several possible summonses and complying with any one
  constitutes success.  We show there is a ``quantum paradox of 
choice'' in summoning: the extra freedom in 
completing the task makes it strictly
harder.  This intriguing result has practical applications for
distributed quantum computing and cryptography and also 
  implications for our understanding of relativistic quantum
  information and its localization in space-time.
\end{abstract}

\maketitle

A Holistic Magician (HM) repeatedly performs the following trick.
He first asks you to give him an object that you are 
sure he cannot copy. 
After working behind a curtain, he presents you with 
$N$ boxes and asks you to choose one. 
Opening your chosen box, he reveals the original object inside. 

You initially imagine that he has arranged some concealed mechanism
that somehow passes the object sequentially through the boxes,
allowing him to stop the mechanism and keep the object in one box if
you select it.  However, you are then puzzled to notice that he is
unable to make the trick work if you select more than one box, even
though you allow him to choose which of your selections to open.  This
argues against your mechanical explanation, and indeed seems to make
any simple explanation problematic.  How can giving the magician more
freedom make him unable to complete the task?  

The so-called paradox
of choice -- more choice can make consumers less happy -- is a
familiar concept in economics \cite{schwartz}.  The magician's
paradox, however, is much sharper: more freedom in choosing how to
execute a task makes it {\it impossible}.  Nonetheless, strange as it
may seem, we show that such a situation can indeed arise when quantum
mechanics is combined with classical relativity.

There is a long-standing tradition of using parables and apparent
paradoxes to refine our understanding of quantum theory \cite{Aharonov,
cat, Cheshire, Zeno, Gibbsparadox, pigeon,lsw}.  
The effect we describe here is perhaps the
first intrinsically relativistic quantum paradox, in the sense that 
it can be formulated only in the framework of relativistic quantum
theory.   It has interesting implications for our theoretical
understanding of relativistic quantum theory and quantum information.
It is also of practical relevance in characterising how quantum states may be propagated in
distributed quantum computers, global financial networks and other
contexts where relativistic signalling constraints are important.

Our paradox involves a task known as summoning, in which an
agent is given an unknown quantum state and required to produce it at
a point in space-time in response to a call made at some earlier point
\cite{Kentnosummoning}. We generalize this task to allow calls to be
made at any number of call points, requiring the agent to return the
state at any return point corresponding to one of the calls.  We
establish necessary and sufficient conditions on the geometric
configuration of call and return points in space-time to guarantee
that the task can be completed.  Further, we show that these are
strictly stronger conditions than those established by Hayden-May (HM) 
for the case where it is guaranteed that at most one call will be
made.   We adopt the approximation that quantum states
may be effectively localized to a point; see Ref. \cite{Hayden} for
further discussion of this approximation and its limitations.    

Summoning was first 
introduced \cite{Kentnosummoning} as a simple illustration of a task 
that distinguishes relativistic quantum theory
from both relativistic classical theories and 
non-relativistic quantum theory.  In its original
form, the task involves two agencies, A and B, 
each comprising collaborating networks of agents
distributed throughout space-time.   B secretly prepares
a random quantum state in some agreed Hilbert
space -- for example, a random qubit -- and gives it
to A at some point $P$.   At some later
point $Q$ in the causal future of $P$, B will ask
for the state back.  The point $Q$ is not known
in advance by $A$.   A simple ``no-summoning
theorem'' shows \cite{Kentnosummoning} that, no matter how densely $A$'s
agents are distributed, and whatever strategy they
use, they cannot generally guarantee to comply with $B$'s request.  
This remains true for variations of the task in which time delays
in returning and some loss of fidelity in the returned state are
allowed \cite{Kentnosummoning}.   

An example of summoning with separated call and return points 
was given in \cite{qtasks}, 
where it was shown that there exist
summoning tasks that cannot be completed by simply propagating the
unknown state along a fixed path but can be completed by the use
of quantum teleportation. 
Hayden-May \cite{Hayden} introduced a generalised version 
of this summoning task defined by a spacetime point $P_s$
and a set of $N$ ordered pairs of spacetime points $\{ c_i, r_i \}$.
Alice is given a quantum system  in some
unknown state $\psi$ at a start point $s$.
If a call is made at point $c_i$, then Alice
must return a quantum system in state $\psi$ at the corresponding
response point $r_i$.   
We follow HM in stipulating that each $r_i$ must be in the
causal future of $c_i$.\footnote{For further discussion of 
this point, see the Supplementary Material}

For definiteness we work in Minkowski space, although our results and most of our
comments apply to more general spacetimes. 
Write $x > y$ if the spacetime point $x$ is in the causal future of
$y$, and $x \geq y$ if either $x>y$ or $x=y$. 
HM define the {\it causal diamond} $D_i$ to be the set
$\{ p : r_i \geq p \geq c_i \}$. 
They then use iterative applications of quantum teleportation and
secret sharing to prove the following beautiful result: 
\begin{theorem} \cite{Hayden} \label{hmtheorem} 
Under the assumptions described, summoning is possible if and only if 
every reveal point $r_i \geq s$ and every pair of causal diamonds $D_i$ and $D_j$ are
causally related, meaning that there exists $x_i \in D_i$ and $x_j \in
D_j$ with $x_i \geq x_j$, or vice versa.  
\end{theorem} 

These are
considerably weaker conditions than naive intuition might suggest.
In particular, there need not necessarily exist a causal path 
that starts from $s$ and runs sequentially through the causal
diamonds.   

Summoning is a natural
task for distributed quantum computing over networks where
relativistic signalling constraints are significant.   One can
imagine quantum data that needs to be routed to one of a number
of nodes, with the destination depending on classical data 
generated other nodes during the
computation.   It is natural for these applications, and also intrinsically
theoretically interesting, to introduce another version of summoning,
in which calls may arrive at any number of the call points.   
Since the $r_i$ may be space-like separated, and since in any
case we assume that the unknown state is handed over to another
agency when returned, the no-cloning theorem means that Alice 
cannot return the state several times.   So we define the task
such that, if several calls are made at points $c_i$ ($i \in I $),
Alice need only return the state at any one of the corresponding
return points $r_i$ ($i \in I$).   

\begin{theorem} \label{newthm} Summoning with unrestricted calls, as
defined above, is possible if and only if (1) every response point $r_i
\geq s$ and (2) for any subset $K$ of $\{ 1 , \ldots , N \}$, there is 
at least one $k \in K$ such that $r_k \geq c_i$ for all $i \in K$.   
\end{theorem} 

The proof is given in the Supplementary Material.  
\vskip 10pt

{\bf Comments} \qquad   {\bf 1.} Any set of causal diamonds
satisfying the conditions of theorem \ref{newthm} also satisfies
those of theorem \ref{hmtheorem}.   However, it is easy to construct examples of sets of causal
diamonds that satisfy those of theorem \ref{hmtheorem} but not those
of theorem \ref{newthm}.   For example, Fig. 1 describes one such set.   Allowing the possibility of more than one call thus makes the
summoning task strictly harder.     

\begin{figure}[t]
\begin{tikzpicture}[scale=0.7]

	%define the call points, the end points, and the start point
	\coordinate (PS) at (2.5,-3.5,1.5);

	\coordinate (Ca) at (0,0,0);
	\coordinate (Cb) at (5,0,0);
	\coordinate (Cc) at (3,0,5);
	
	\coordinate (Qa) at (2.5,2.5,0);
	\coordinate (Qb) at (4, 2.5, 2.5);
	\coordinate (Qc) at (1.5, 2.5, 2.5);
	
	%draw the frame
	\draw[dashed] (Ca) -- (Cb) -- (Cc) -- (Ca);
	\draw[dashed] (0,2.5,0) -- (5,2.5,0) -- (3,2.5,5) -- (0,2.5,0);
	\draw[dashed] (0,2.5,0) -- (0,0,0);
	\draw[dashed] (5,2.5,0) -- (5,0,0);
	\draw[dashed] (3,2.5,5) -- (3,0,5);
	
	%draw the regions
	\draw[ultra thick] (Ca) -- (Qa);
	\draw[ultra thick] (Cb) -- (Qb);
	\draw[ultra thick] (Cc) -- (Qc);
	
	%draw the causal connections
	\draw[-triangle 45][red] (Ca) -- (Qc);
	\draw[-triangle 45][red] (Cb) -- (Qa);
	\draw[-triangle 45][red] (Cc) -- (Qb);
	
	%draw the end points
	\draw[fill=blue] (Qa) circle (0.15);
	\draw[fill=blue]  (Qb) circle (0.15);
	\draw[fill=blue]  (Qc) circle (0.15);
	
	\node [above right] at (Qa) {$r_1$};
	\node [above left] at (Qb) {$r_2$};
	\node [above right] at (Qc) {$r_3$};
	
	%draw the call points
	\draw[fill=black] (Ca) circle (0.15);
	\draw[fill=black] (Cb) circle (0.15);
	\draw[fill=black] (Cc) circle (0.15);
	
	\node [left] at (Ca) {$c_1$};
	\node [below right] at (Cb) {$c_2$};
	\node [below right] at (Cc) {$c_3$};
	
	%draw the start point
	\draw[fill=yellow] (PS) circle (0.15);
	\node [right] at (PS) {$s$};
	
	%draw some axes
	\draw[->] (-0.5,-3.5,1.5) -> (0.5,-3.5,1.5);
	\draw[->] (-0.5,-3.5,1.5) -> (-0.3,-3.5,0.3);
	\draw[->] (-0.5,-3.5,1.5) -> (-0.5,-2.5,1.5);
	
	\node[below] at (0.5,-3.5,1.5) {x};
	\node[right] at (-0.3,-3.5,0.3) {y};
	\node[left] at (-0.5,-2.5,1.5) {t};
	
\end{tikzpicture}
\caption{A $2+1$ dimensional example, taken from Ref \cite{Hayden}, where summoning is
  possible if one call is guaranteed, but not if more than one call
may arrive.   The black lines are lightlike and represent singular
causal diamonds; the red arrows are also lightlike. 
}\label{fig:qec}
\end{figure}
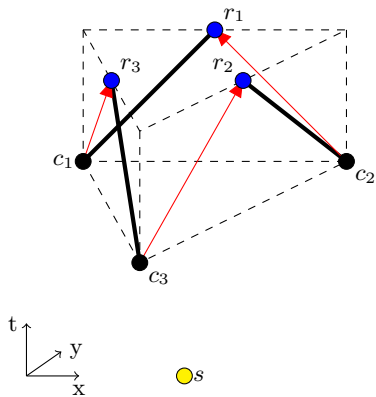

{\bf 2.}   Nonetheless, the conditions of theorem \ref{newthm}
still do not imply that there is a causal path running from the start point
through each causal diamond.    
An example is given in Fig. 2.  
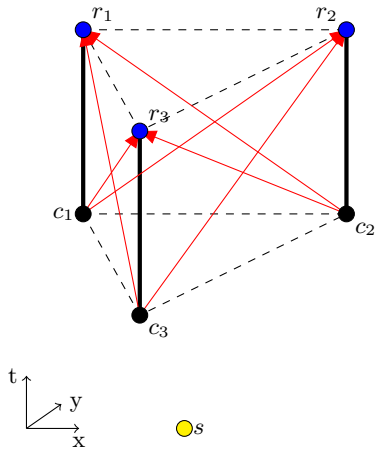
\begin{figure}[t]
\begin{tikzpicture}[scale=0.7]

	%define the call points, the end points, and the start point
	\coordinate (PS) at (2.5,-3.5,1.5);

	\coordinate (Ca) at (0,0,0);
	\coordinate (Cb) at (5,0,0);
	\coordinate (Cc) at (3,0,5);
	
	\coordinate (Qa) at (0,3.5,0);
	\coordinate (Qb) at (5, 3.5, 0);
	\coordinate (Qc) at (3, 3.5, 5);
	
	%draw the frame
	\draw[dashed] (0,3.5,0) -- (5,3.5,0) -- (3,3.5,5) --
        (0,3.5,0);
	\draw[dashed] (0,0,0) -- (5,0,0) -- (3,0,5) -- (0,0,0);
	
	%draw the regions
	\draw[ultra thick] (Ca) -- (Qa);
	\draw[ultra thick] (Cb) -- (Qb);
	\draw[ultra thick] (Cc) -- (Qc);
	
	%draw the causal connections
	\draw[-triangle 45][red] (Ca) -- (Qc);
	\draw[-triangle 45][red] (Ca) -- (Qb);
	\draw[-triangle 45][red] (Cb) -- (Qa);
	\draw[-triangle 45][red] (Cb) -- (Qc);
	\draw[-triangle 45][red] (Cc) -- (Qa);
	\draw[-triangle 45][red] (Cc) -- (Qb);

	%draw the end points
	\draw[fill=blue] (Qa) circle (0.15);
	\draw[fill=blue]  (Qb) circle (0.15);
	\draw[fill=blue]  (Qc) circle (0.15);
	
	\node [above right] at (Qa) {$r_1$};
	\node [above left] at (Qb) {$r_2$};
	\node [above right] at (Qc) {$r_3$};
	
	%draw the call points
	\draw[fill=black] (Ca) circle (0.15);
	\draw[fill=black] (Cb) circle (0.15);
	\draw[fill=black] (Cc) circle (0.15);
	
	\node [left] at (Ca) {$c_1$};
	\node [below right] at (Cb) {$c_2$};
	\node [below right] at (Cc) {$c_3$};
	
	%draw the start point
	\draw[fill=yellow] (PS) circle (0.15);
	\node [right] at (PS) {$s$};
	
	%draw some axes
	\draw[->] (-0.5,-3.5,1.5) -> (0.5,-3.5,1.5);
	\draw[->] (-0.5,-3.5,1.5) -> (-0.3,-3.5,0.3);
	\draw[->] (-0.5,-3.5,1.5) -> (-0.5,-2.5,1.5);
	
	\node[below] at (0.5,-3.5,1.5) {x};
	\node[right] at (-0.3,-3.5,0.3) {y};
	\node[left] at (-0.5,-2.5,1.5) {t};
\end{tikzpicture}

\caption{An example with $c_1 , c_2 , c_3 <  r_1 , r_2 , r_3$.  The diamonds satisfy the conditions of
theorem \ref{newthm}, and so summoning with any number of calls is possible, even though there is no causal path running through all
three diamonds.    The timelike black lines run between the call
points and their return points and define the centres of the diamonds
at each time; the red arrows are lightlike. }
\end{figure}

{\bf 3.}  The conditions of theorem \ref{newthm} imply there is an 
ordered sequence of causal diamonds $D_{i_1} , \ldots , D_{i_N}$ 
such that the return point of any diamond in the sequence lies in the 
causal future of the call points of all previous diamonds.  This 
ordering is not necessarily unique.   For example, a nested pair of
diamonds, with $c_i \leq c_j < r_j \leq r_i $ with appropriate relations 
to the other diamonds may be taken in either order.   
More generally, one can construct examples including sets of 
$n$ non-overlapping diamonds $(c_i , r_i )$ (for $1 \leq i \leq n$)
for which $c_i < r_j$ for all $i,j$; Fig. 2 gives an example of this
type for $n=3$.     
%A simple such example with
%$n=2$ is given by singular (lightlike line segment) ``diamonds'' 
%with endpoints $(c_1, r_1 ) = ( ( -1,0,-1), (1,0,1)), 
%(c_2 , r_2 ) = ((1,y,-1),(-1,y,1)) $ for small $y$.  

{\bf Resolution}  \qquad The resolution of the apparent paradox rests
on a previously unappreciated feature of summoning tasks. 
Prima facie it seems that the guarantee of at most one 
call plays no special role in a summoning task other than to ensure
that Alice is never required to produce two copies of an unknown
state, in violation of the no-cloning theorem.
It thus initially seems paradoxical that summoning becomes strictly
harder if we allow the possibility of more than one call, even though
only one valid response is required.   

However, if Alice knows that no more than one call will occur,
learning that a call has been made at one point tells
her that there are no calls at any other point, and this allows her to coordinate
the behaviour of her agents via the global call distribution.   
A single call gives Alice less information if multiple calls can occur:
she learns nothing about the distribution of calls at other points. 
She thus cannot use the call distribution to coordinate her  
agents' actions in the same way. 
In other words, the guarantee of at most one call provides a resource
that gives Alice the ability to complete tasks that would be impossible without
it. 

{\bf Applications} \qquad We have given necessary and sufficient conditions for an algorithm
that may call for a state at several distinct nodes but requires the
state to be produced at only one associated response node.  This is a
natural condition in many practical contexts.  For example, one can
imagine a distributed parallel quantum computation in which the output
of a sub-protocol should be routed to one of several parallel
computations, which call for the output when they reach a certain
state.  Indeed, in the teleportation model of distributed quantum computation
\cite{Gottesman}, each round of adaptive computation is
essentially a summoning task: the measurement result from the previous
round determines the measurement to be made in the present round, and
thus plays the role of the `call', while the locations of the gates
for the various possible measurements play the role of the `response
points'.

Our conditions are stronger than those required if it is known
there will be at most one call, but substantially weaker than
those required if the state were to propagate along some definite
causal path.   The no-summoning theorem has also already led to new 
applications in relativistic quantum cryptography \cite{bcsummoning,
bcmeasurement,lunghietal,AdlamKent1,AdlamKent2}.  The 
stronger results reported here and in Ref. \cite{Hayden} 
suggest further ways of exploiting summoning as a 
general way of controlling the flow of quantum information.   We 
thus expect these results
to find application in future cryptographic protocols as well as in
quantum network algorithms.  

Our results also have intriguing implications regarding the nature
of relativistic quantum information and its spatio-temporal
localization.  We discuss these further in the Supplementary Material.  

%\bibliographystyle{apsrev4-1}

%\bibliography{qpoc5}{}
%merlin.mbs apsrev4-1.bst 2010-07-25 4.21a (PWD, AO, DPC) hacked
%Control: key (0)
%Control: author (0) dotless jnrlst
%Control: editor formatted (1) identically to author
%Control: production of article title (0) allowed
%Control: page (1) range
%Control: year (0) verbatim
%Control: production of eprint (0) enabled
%

\end{document}

\section{Supplementary Material}

\subsection{Proof of Theorem 2}

\textbf{Necessity:} The no-signalling principle implies that the first condition is
necessary.  
To see that the second condition is necessary, suppose for the purpose
of obtaining a contradiction that there exists a successful protocol
for any subtask defined from the original task by selecting some set
$\mathcal{M}$ of $M \leq N$ call-response pairs, such that for every
call-response pair $(c_i, r_i) \in \mathcal{M}$, there exists at least
one call-response pair $(c_j, r_j) \in \mathcal{M}$ such that $r_i
\ngtr c_j$.   

Since each return point is in the casual future of the corresponding
call point, Alice may decline to return anything at $r_i$ if no call is made at
$c_i$.  We can thus assume Alice's strategy returns no state anywhere if no
calls are made.  

From the no-signalling principle, the response made at $r_i$ when
calls are made at some set of call points $Q $ with $c_i, c_j \in Q$
is the same as the response made at $r_i$ when calls are made at the
set $\{ Q \setminus c_j \}$.

Therefore for any $r_i$, if $\mathcal{Q}_i$ is the total number of
subsets $Q \subseteq \{ c_1, c_2, ... c_M \}$ such that $\psi$ is
returned at $r_i$ if calls are made at all points in $Q$, then
$\mathcal{Q}_i$ must be even.
Therefore $\sum_{i = 1}^M \mathcal{Q}_i$ must be even. 

Since Alice must respond at exactly one point whenever calls are made
at any subset $Q \subseteq \{ c_1, c_2, ... c_M \}$, we must have
$\sum_{i = 1}^M \mathcal{Q}_i = \sum_{j = 1}^M C^M_j 
= 2^M - 1$, which is always odd. 

So there exists a successful protocol for the sub-task defined by the
set $\mathcal{M}$ only if there exists at least one call-response pair
$(c_i, r_i) \in \mathcal{M}$ such that $r_i$ is in the future
lightcone of the call point for every pair in $\mathcal{M}$.

Since this reasoning may be applied to any sub-task, there exists a
successful protocol for a summoning task without guarantee with $N$
call-response pairs only if for any subset $S$ of call-response pairs,
at least one response point $r_k $ belonging to a pair in $S$ lies in
the future lightcone of all the other call points belonging to pairs
in $S$.
\textbf{Sufficiency:}
We now show that the conditions are sufficient, by exhibiting a
protocol that
always succeeds if the call-response pairs satisfy the conditions of
theorem \ref{newthm}. 
Define $S_N = \{ 1 , \ldots , N \}$. 
From the second condition of theorem \ref{newthm}, there is at least one $i \in S_N$ 
such that $r_i \geq c_j$ for all $j \in
S_N$.  Choose one such, $i_N$, and define $S_{N-1} = S_N \setminus \{ i
\}$.   Similarly, choose $i_{N-1} \in S_{N-1}$ such that $r_{i_{N-1}}
\geq c_j$ for all $j \in
S_{N-1}$, and so on.  We thus obtain an 
ordered sequence of causal diamonds $D_{i_1} , \ldots , D_{i_N}$ 
such that the return point of any diamond in the sequence lies in the 
causal future of the call points of all previous diamonds. 
We relabel the $D_{i_j} , c_{i_j}, r_{i_j}$, writing $j$ for $i_j$.  

Alice may now proceed as follows.   Let $d$ be the agreed dimension
of the Hilbert space of the unknown state $\psi$. Before the protocol begins,
she distributes maximally entangled pairs of states in $\mathcal{C}^d \otimes
\mathcal{C}^d$ between agents at the spatial locations of $(s,c_1), 
(c_1 , c_2), \ldots, (c_{N-1}, c_N )$.      
When the state is given to her at $s$, her agent there teleports it to $c_1$, 
broadcasting the classical teleportation data.   
If a call arrives
at $c_1$, the agent at $c_1$ transmits the quantum part of the teleported state
(i.e. her half of the entangled pair) to $r_1$, where another 
agent uses the classical teleportation data to reconstruct the state
and return it to Bob.    
 
If a call does not arrive at $c_1$, Alice's agent there teleports the
state to $c_2$, broadcasting the classical teleportation data. 
If a call arrives at $c_2$, the quantum part of the teleported state
is sent to $r_2$, where the classical teleportation data from $s$ and 
$c_1$ are used to reconstruct and return the state. 
Otherwise, the process continues, until either a call arrives at 
some $c_i$ (and the state is reconstructed and returned at $r_i$) 
or there is no call.   In the latter case, Alice may 
reconstruct the state at $r_N$ if she wishes, but does not return it to Bob.

\subsection{Summoning without restrictions on the geometry} 

In the main text, we follow HM in assuming that each return point
$r_i$ lies in the causal future of its call point $c_i$. 
Actually, this assumption is slightly restrictive, since it can be 
possible to complete the task even when one of the $r_i$ lies
outside the causal future of its call point $c_i$.  However, the 
theorem has a simple generalisation that covers this case, as we show
below.  

A related issue is that HM stipulate that at most one call will be made, but do not explicitly 
define the task requirements if no call is made.   
One natural definition is to require that Alice should return a quantum state
at $r_i$ {\it only if} she is certain at $r_i$ that a call was made at $c_i$.
Hence, she should not return any state anywhere if no call is made. 
With this refinement, it follows that each return point must indeed
be in the future of the corresponding call point, as HM assume.

However, another natural definition of the task is to allow Alice to
return a quantum state anywhere or nowhere, as she chooses, if no
call is made.   In one version of this task, any returned state
must be the state originally supplied, even if no call is made.
Another possibility is to allow Alice to return any state when
no call is made.  Even the first, stronger, version, is possible
for configurations where theorem \ref{hmtheorem} does
not apply, as we now show.  

Recall again that the conditions of theorem
\ref{hmtheorem} assume that each $r_i$ lies in the causal future
of its call point $c_i$, so that each causal diamond $D_i$ is 
well-defined and non-empty.    The theorem then shows that summoning is possible
only if for every pair $i, j$, the causal diamonds $D_i$ and $D_j$ are
causally related.    

Consider now the case where we have only two possible call points,
$c_1$ and $c_2$, with corresponding return points $r_1$ and $r_2$. 
If summoning is possible, $r_1$ and $r_2$ must lie in the future of
the start point $s$.   However, summoning may be possible even if one
of the causal diamonds $D_1$, $D_2$ is empty, i.e. even if one of the
$r_i$ does not lie in the future of $c_i$. 

This is because in the case of only two call points, learning that no
call is made at $c_1$ implies that either there is a call at $c_2$ or no call at all.
Hence, if $r_2 > c_1$,  we can construct a successful protocol even
if it is not the case that $r_2 > c_2$.
\begin{enumerate} 

\item Before the protocol starts, Alice shares an entangled pair
between agents who will be at $s$ and $c_1$.  

\item When the state is handed over at $s$, Alice teleports it to
  $c_1$, broadcasting the classical data in all directions.  

\item If a call is received at $c_1$, Alice's agent at $c_1$ sends the
  quantum teleportation data (i.e. her part of the entangled pair) to
  $r_1$, where it is combined with the classical teleportation data to reconstruct
  the state, which is handed over to Bob.   Alice's agent at $c_1$
  broadcasts the fact that she has
received a call.  Alice's agent at $r_2$ then does not hand over anything.   

\item If no call is received at $c_1$, Alice's agent there sends the
quantum teleportation data to $r_2$, where it is combined with the
classical teleportation data and handed over to Bob.  

\end{enumerate} 

Thus if a call is made at $c_1$ the state is at $r_1$ at the end of
the protocol; if a call is made at $c_2$ (or no call is made at all)
the state is at $r_2$ at the end of the protocol. 

Hayden-May's proof \cite{Hayden} shows that a protocol for
any number of call points can be built recursively out of a protocol for two call
points.   Their argument extends to the more general configurations
we consider.    
Theorem \ref{hmtheorem} can thus be extended as follows:

\begin{theorem} A summoning task defined by a set of $N$ ordered pairs $\{ c_i, r_i \}$ is possible iff the following conditions hold: 

\item Every response point is in the future lightcone of the start point. 

\item For every pair $\{c_i, r_i \}, \{c_j, r_j \}$, both of the response points lie in the future lightcone of a single call point. 

\end{theorem} 

Note that it follows that for every pair $i, j$ at least one of the
causal diamonds $D_i$, $D_j$ must exist, so there can be at most one
call-response point pair for which the causal diamond does not exist.

\subsection{Discussion and interpretation}

Like the original no-summoning theorem, our results rely
crucially on both relativity and quantum theory.  
In particular, the distinction between summoning tasks with and without guarantees
of a single call, and indeed that there are interesting
constraints on summoning tasks at all, relies both on the 
impossibility of superluminal signalling and on the
no-cloning theorem.   Our results also exploit the delocalizability of
quantum information via quantum teleportation, which allows summoning
in configurations where sending the state along any given causal path fails.

Our results thus allow us to probe
intuitions about the nature of quantum states as spatiotemporal
entities -- a question which has received comparatively little attention
in recent debates over the reality of the quantum state
\cite{HarriganSpekkens,PBR,Quanta22,Hardy2013,RennerColbeck, cr1, Maroney,
  Maroney2, Montina}.
There is a tradition in physics, thus far largely unanalysed, of describing
quantum states and quantum information using the language
of persisting physical objects \cite{Timpson}.  For example, 
Horodecki et al.'s widely cited review \cite{Horodeckientanglement} states: 
``This is the essence of
teleportation: a quantum state is transferred
from one place to another: not copied to other place, but
moved to that place.'' 
Hayden-May follow in this tradition in interpreting their result: ``We fully characterize which regions of
spacetime can all hold the same quantum information. Because quantum
information can be delocalized through quantum error correction and
teleportation, it need not follow well-defined trajectories.  Instead,
replication of the information in any configuration of spacetime
regions not leading to violations of causality or the no-cloning
principle is allowed. $\ldots$ This provides a simple and complete
description of where and when a qubit can be located in spacetime,
revealing a remarkable variety of possibilities.''\cite{Hayden}

We question whether HM's spatiotemporal account of the
quantum state during summoning tasks can be supported. 
Indeed, even before considering the implications of theorems
\ref{hmtheorem} and \ref{newthm}, we find it a little tricky to
understand what exactly Hayden-May's quotation is intended to mean.

One technical complication is that the causal diamonds in a summoning
task may overlap or even be identical.   In this case, even if two overlapping diamonds in
some sense each hold a copy of the state, it could be a single copy held in their 
intersection.  HM do not explicitly consider the possibility of
overlaps.  Since we wish to consider the case most favourable to 
their interpretation, we assume non-overlapping diamonds here.   

A more fundamental concern is that, if the quantum information 
were replicated in the usual sense within each causal diamond in a
configuration satisfying HM's conditions, ordinary reasoning would
suggest that, given a request at any number of call points,
it should be possible to produce a copy of the state 
at each of the corresponding return points.  This would violate the
no-cloning theorem, of course, and evidently is not the intended
meaning -- hence HM's phrase {\it ``replicated but in a restricted fashion''}. 
We are clearly meant somehow to allow
for the fact that replication in time is not the same as replication
in space -- but how, exactly?   

HM's comment that
``reversibility $\ldots$ requires that [quantum] information be
copied in time'' suggests the following analogy.  The  
quantum information in an unknown state with known unitary evolution
can be captured and handed over at any given time, or more generally on any given spacelike
hypersurface in a foliation.  Once this happens, though, it cannot be
repeated: if the state is requested again at later times
(or on later hypersurfaces), Alice cannot provide a second copy.   
An analogous statement about summoning would 
be that, when the conditions of theorem \ref{hmtheorem} hold,
if several requests are made, the quantum information
can be handed over in one causal diamond, but then this cannot be repeated in the other diamonds.  
However, as our result shows, this is not generally the case!   
Requesting the state at several HM call points {\it prevents} a successful 
response, {\it unless} the configuration of call and return points also satisfies the stronger conditions
of our theorem \ref{newthm}.   

Unlike the conditions of theorem \ref{hmtheorem}, those of theorem \ref{newthm} allow an
{\it ordering} of the diamonds, albeit not necessarily a unique one. 
Moreover, they allow a strategy that effectively (although
not literally) responds to the calls in the chosen ordered sequence.  
We believe this is a crucial distinction. 
It takes us closer to the case of a
unitarily evolving state that can be requested at any time or on any
hypersurface in a foliation, for which the time coordinates or
hypersurfaces are likewise ordered.  And indeed, {\it if} Alice
follows our strategy, then at each call point $c_i$ she knows that, {\it if} the state is
not being returned as the result of an earlier call, and if she
chooses not to teleport her state to $c_{i+1}$, then -- at least in
the informal language physicists tend to use when discussing
teleportation -- the unknown state's quantum information {\it will} be
contained within $D_i$.  That is, there is a quantum state at $c_i$,
together with classical information available at or before $r_i$,
which could be brought together and recombined to reconstruct the
state at $r_i$.

So what {\it can} we consistently say about the flow of quantum information in
summoning tasks, in the light of our results?  One safe option, of
course, is to remain silent, retreating instead to a purely
operationalist perspective.  Proving results about possible and
impossible quantum tasks in space-time \cite{qtasks} does not
necessarily require any discussion of the localization or location of
quantum information in space-time, as theorems \ref{hmtheorem} and
\ref{newthm} illustrate.   Operationalism avoids the risk of 
fallacies that can arise from regarding (quantum) information as a 
physical substance \cite{Timpson}.

That said, whatever view one takes on the reality of quantum
states, even speaking informally about the localization of 
quantum information could be a very helpful way of 
proving and synthesizing operational results,
if it accurately reflects their implications. 
Our suggestion is thus not that such language should necessarily be 
eliminated altogether, but rather that it needs to be analysed
critically and used more precisely.   
In particular, if the behaviour of quantum information in a system 
depends on external events, such as the calls in a summoning task,
we would argue that any discussion of its localization should reflect this explicitly.
For example, a more precise informal statement of Hayden-May's result would
be that {\it if} Alice follows their prescribed strategy, and {\it if} a call
arrives at $c_i$ and no call arrives at any other call point, {\it then}
the unknown state's quantum information becomes localized within $D_i$
and the state is reconstructed at $r_i$.   This framing precludes
speaking of ``replication of the information in [the given]
configuration of spacetime regions'', since the possibilities of 
localizing the state within $D_i$ and $D_j$ are alternatives 
that depend on exclusive possible external events.   
We suggest too that any statements about the localization of quantum
information in space-time -- including 
those we give above for our version of the task -- may be best  
seen as hypotheses that seem to usefully synthesize and summarize operational results,
but might yet need to be refined or even rejected.

\subsection{Further comments on the
quantum state and spatiotemporal reality} 

Our results directly apply only to summoning tasks; they do not 
give a definitive answer to the general question about
when and whether it is correct (or at least consistent) to 
think of quantum states as though they were real physical
objects that can move in space and persist over time (albeit
perhaps discontinuously, as the quotation from
Ref. \cite{Horodeckientanglement} suggests is the case
during teleportation).   Still, they suggest to us 
that this and related questions deserve closer analysis
and that examining relativistic quantum tasks \cite{qtasks} might
yield further insights.       

To take one interesting example, the celebrated PBR theorem \cite{PBR,
  Quanta22} purportedly addresses
only the question of whether the quantum state is an element of
reality in the instantaneous sense set out by Harrigan and Spekkens
\cite{HarriganSpekkens}. 
It seems to us that some caution is merited here, since the proof
of the theorem depends implicitly on assumptions not only about states
at a given time, but about the persistence of those states over
time. Specifically, the argument hinges on the assumption of
preparation independence, the idea that `systems that are prepared
independently have independent physical states,' and it is assumed
without comment that the systems must therefore continue to have
independent physical states when they are brought together and
measured, in which case it is possible to derive a contradiction
between the predictions of quantum mechanics and the predictions of
any model which allows some overlap between the ontic states
associated with different quantum states. Several authors
\cite{Emersonetc, Quanta22, ABCL} 
have pointed out that the argument
fails if we relax the assumption of preparation independence.
However, these 
criticisms of preparation independence have been directed almost
exclusively at the claim that the states are independent at the time
of preparation.  For example, Ref. \cite{Emersonetc} proposes that even
when a systems are prepared in a product state they might have global
properties not reducible to properties of their subsystems, while
Ref. \cite{Hall} suggests the preparations might fail to be independent due
to their common past.  Less attention has been given to the fact that PBR
must also claim that these independent states persist over time so
that the systems still have the same independent states at the time of
the measurement.   It seems to us that what the PBR theorem really
establishes is that \emph{if} quantum systems have localized physical 
states which persist over time, and the other explicit assumptions
of the theorem hold, then those states must carry all the same 
information as the quantum state. 
The theorem thus carries no weight against views which deny this
spatiotemporal account of physical states in the first place.
For example, the theorem {\it per se} offers no reason to reject 
any view which suggests that measurement results may depend
directly on preparations without being mediated by any persisting
physical state, quantum or otherwise.  

It is true that there exist later $\psi$-ontology theorems which come closer to
addressing the kind of spatioemporal concerns we raise here.
In particular, Hardy's theorem \cite{Hardy2013} and the Colbeck-Renner
theorem \cite{RennerColbeck} both concern themselves with dynamics
rather than merely kinematics, and so deal more explicitly with the
passage of time. However these theorems likewise work within the
established ontological models framework which presupposes that past
events may influence future ones only via the mediation of some
persisting physical state, and thus like PBR, can be circumvented by
denying this sort of spatiotemporal account of quantum states. 
This suggests to us that a more detailed analysis of possible and
impossible relativistic quantum tasks, and of their implications for
hypotheses about the temporal persistence of quantum and ontic states,
might shed some interesting new light on our understanding of 
the quantum state and its relationship to reality.  

%\bibliography{qpoc5}{}
%merlin.mbs apsrev4-1.bst 2010-07-25 4.21a (PWD, AO, DPC) hacked
%Control: key (0)
%Control: author (0) dotless jnrlst
%Control: editor formatted (1) identically to author
%Control: production of article title (0) allowed
%Control: page (1) range
%Control: year (0) verbatim
%Control: production of eprint (0) enabled
%